\documentclass[twocolumn, doublespace, 12pt fonts]{aastex6}
\usepackage{amsmath}
\usepackage{graphicx}
\usepackage{bm}
\usepackage{epsf}

\shorttitle{Pulsar mass beyond $\sim 2.3~M_{\odot}$} 
\shortauthors{Wu, Du, \& Xu}
\begin{document}

\title{What if the neutron star maximum mass is beyond $\sim2.3 M_{\odot}$?}

\author{Xuhao Wu, Shuang Du and Renxin Xu}
\affil{School of Physics, Peking University, Beijing 100871, China
}%
\email{x.h.wu@pku.edu.cn,\\
r.x.xu@pku.edu.cn}
 
\begin{abstract}
By assuming the formation of a black hole soon after the merger event of GW170817, Shibata et al. 
updated the constraints on the maximum mass ($M_\textrm{max}$) of a stable neutron star within $\lesssim$ 2.3 $M_{\odot}$, but there is no solid evidence to rule out $M_\textrm{max}>2.3~M_{\odot}$ from the point of both microphysical and astrophysical views.
In order to explain massive pulsars, it is naturally expected that the equation of state (EOS) would
become stiffer beyond a specific density.
In this paper, we consider the possibility of EOSs with $M_\textrm{max}>2.3~M_{\odot}$, 
investigating the stiffness and the transition density in a polytropic model.
Two kinds of neutron stars are considered, i.e., normal neutron stars (the density vanishes on gravity-bound surface) and strange stars (a sharp density discontinuity on self-bound surface).
The polytropic model has only two parameter inputs in both cases: ($\rho_{\rm t}$,
$\gamma$) for gravity-bound objects, while ($\rho_{\rm s}$, $\gamma$) for
self-bound ones, with $\rho_{\rm t}$ the transition density, $\rho_{\rm s}$ the
surface density and $\gamma$ the polytropic exponent.
In the matter of $M_\textrm{max}>2.3~M_{\odot}$, it is found that the smallest $\rho_{\rm t}$ and $\gamma$ should be $\sim 0.50~\rho_0$ and $\sim 2.65$ for normal neutron stars, respectively, 
whereas for strange star, we have $\gamma > 1.40$ if $\rho_{\rm s} > 1.0~\rho_0$ and $\rho_{\rm s} < 1.58~\rho_0$ if $\gamma <2.0$ ($\rho_0$ is the nuclear saturation density).
These parametric results could guide further research of the real EOS with any foundation of microphysics if a pulsar mass higher than $2.3~M_{\odot}$ is measured in the future.
We also derive rough results of common neutron star radius range, which is $9.8~\rm{km} < R_{1.4} < 13.8~\rm{km}$ for normal neutron stars and $10.5~\rm{km} < R_{1.4} < 12.5~\rm{km}$ for strange stars.
\end{abstract}

\keywords{equation of state - stars: neutron}

\section{Introduction}
The equation of state (EOS) of dense matter, especially of ultra-dense matter, 
is a key issue in nuclear physics and astrophysics (Weber\citealt{Weber2005}).
There are two kinds of neutron star scenarios, gravity-bound system, and self-bound system. The conventional neutron star is a gravity-bound system, with gravity-bound surface, usually has smaller radius with larger mass. In contrast, strange star (strange quark star (Alcock, et al.\citealt{Alcock1986}) and strangeon star (Lai \& Xu\citealt{Lai2017})) as self-bound system, that their surface are self-bounded,  has larger radius with larger mass.
The normal neutron star is divided into the core part and the crust part. Physicists developed many-body theories to describe the core and inner crust EOS, which is unclear at high density, such as Green Function Monto Carlo (GFMC)
method (Pieper \& Wiringa\citealt{Pieper2001}), chiral perturbation theory (ChPT) (Gasser \& Leutwyler\citealt{Gasser1984}), Brueckner-Hartree-Fock (BHF) (Brockmann \& Machleidt\citealt{Brockmann1990}), quark mean-field (QMF) model (Toki et al.\citealt{Toki1998}), quark meson coupling (QMC) model (Guichon\citealt{Guichon1988}, Saito \& Thomas\citealt{Saito1994}), relativistic mean-field (RMF) model (Serot \& Walecka\citealt{Serot1986})  et al.
The Baym-Pethick-Sutherland (BPS) (Baym et al.\citealt{Baym1971}) EOS is commonly used to describe the neutron star out crust (lower than neutron drip density).

Neutron star merger event GW170817 offers a limit of tidal deformability, $70 \le \Lambda_{1.4} \le 580$ (Abbott, et al.\citealt{Abbott2017},\citealt{Abbott2018}).  Based on various many-body methods and this tidal deformability range, a roughly consistent $1.4~M_{\odot}$ neutron star radius constraint refers $R_{1.4} \leq 13.6$ km (Annala, et al.\citealt{Annala2018}, Krastev \& Li \citealt{Krastev2019}, Tews\citealt{Tews2018a}) using the original findings $\Lambda_{1.4} \le 800$ (Abbott, et al.\citealt{Abbott2017}).
Based on NASA's Neutron Star Interior Composition Explorer (NICER; Gendreau et al.\citealt{Gendreau2016}) data set, it is able to estimate neutron star mass and radii using X-ray pulse-profile modeling 
(Raaijmakers et al.\citealt{Raaijmakers2019}).
Neutron star radii as a observable quantity is valuable to restrict the EOS.
We conclude this $R_{1.4} \leq 13.6$ restrict in our paper as a contrast of the tidal deformability constraint.
It is believed the merger event may form a transitory state like hypermassive or supermassive neutron star and eventually becomes black hole.  Based on this assumption, Shibata et al.\cite{Shibata2019} employing the energy and angular momentum conservation laws and numerical-relativity simulations get the cold spherical neutron star maximum mass $M_{\rm{max}} \lesssim 2.3~M_{\odot}$. 
However the $M_{\rm{max}} \lesssim 2.3~M_{\odot}$ constraint may be exceeded if the final remain is a stable supermassive neutron star rather than a black hole.
The 2 $M_{\odot}$ pulsar observations (PSR J0348+0432~Antoniadis, et al.\citealt{Antoniadis2013}, PSR J1614-2230~Demorest, et al.\citealt{Demorest2010}, Fonseca, et al.\citealt{Fonseca2016}, J0740+6620~Cromartie et al.\citealt{Cromartie2020}) and $R_{1.4} \leq 13.6$ km constraints have ruled out many EOSs (Zhu et al.\citealt{Zhu2018}). 
Usually stiff EOSs lead to high maximum mass and large radii while soft EOSs correspond to low maximum mass and small radii.
For the normal neutron star scene, the appearance of new degrees such as quark, meson condensation, hyperon, $\Delta$ particles always soften the EOS, but the quarkyonic matter assumption makes it possible to have a stiffer EOS (Fukushima \& Kojo\citealt{Fukushima2016}, McLerran \& Reddy\citealt{McLerran2019}). Hadron-quark crossover phase transition usually has a hard core that the quarkyonic matter EOS is stiffer than hadronic matter EOS.
Haensel et al.\cite{Haensel1981} studied the saturation density effect on the mass-radius relation, which implies smaller saturation density could support higher maximum mass. The EOS stiffness is usually measured by sound of speed or adiabatic index (Tews et al.\citealt{Tews2018}, Potekhin et al.\citealt{Potekhin2013}) and a stiffer EOS is always required for massive pulsar.
Conventional EOSs have several parameters, but a single 2-parameter family could offer an accessible approximation (Ofengeim\citealt{Ofengeim2020}).
Polytropic model (Chandrasekhar\citealt{Chandrasekhar1939}, Raithel et al.\citealt{Raithel2016}) has only two parameters (the polytropic constant, $K$ and the polytropic exponent $\gamma$) and could model normal phase or exotic phase (Lai \& Xu\citealt{Lai2009}).
Baron et al.\cite{Baron1985}  use the EOS combining the compressible liquid-drop model EOS (Cooperstein\citealt{Cooperstein1985}) and a polytropic model which use incompressibility as the pressure coefficient. They obtained relative small maximum mass due to small $K$-parameter.
We apply the polytropic model on these gravity-bound and self-bound scenarios, and focus on the mass-radius relation which offers strong constraints.  
The results show that a smaller transition density $\rho_{\rm{t}}$ (or surface density $\rho_{\rm{s}}$) and a larger polytrope of exponent $\gamma$ are always beneficial  for stiffer EOS then larger maximum mass. For the neutron star scene, the $R_{1.4}$ limits the transition density which can not be too small. 
For normal neutron star of $M_{\rm{max}} >2.3~M_\odot$ , the smallest transition density and polytrope of exponent  are ($\rho_{\rm{t}}/\rho_0, \gamma$)=(0.50, 2.65). 
While for strange star of $M_{\rm{max}} >2.3~M_\odot$ , the $\rho_{\rm{s}}/\rho_0$=(1.0 $\sim$ 2.0) correspond polytropic exponent region should be $\gamma>1.40$ if $\rho_{\rm{s}}/\rho_0>1.0$, and also $\rho_{\rm{s}}/\rho_0<1.58$ when $\gamma < 2.0$. A smaller polytropic exponent in the self-bound system could derive similar maximum mass than in gravity-bound system. We give details in the results part  for reference.

In this paper, we attend to use the simple polytropic model to clarify how stiff the pulsar EOS should be and where the stiff EOS  starts. In the next section we describe the polytropic model and discuss the selection of the parameters. The gravity-bound and self-bound scenarios are discussed with two free parameters. In section 3, we compare the EOS stiffness and mass-radius constraints in the parameter space. There is a conclusion at last.
\section{The Models}
We apply two kinds of pulsar scenarios, gravity-bound star and self-bound star. Normal neutron star is a gravity-bound system, which usually has smaller radius with larger mass. In contrast, strange star as the self-bound system has larger radius with larger mass.
\subsection {Gravity-Bound Object on Surface}
We assume the neutron star has a soft crust and a hard core to support higher mass.
Our EOS is a combination of BPS EOS and a polytropic model. We apply the BPS EOS when $\rho<\rho_{\rm{drip}}$, with $\rho_{\rm{drip}}$ the neutron drip out density. We assume a transition density $\rho_{\rm{t}}$, after which the EOS becomes stiff to support a massive core. When $\rho \leq \rho_{\rm{t}}$, the pressure is assumed as the extension of the BPS EOS,
\begin{equation}
P_{1}\left(\rho \right)=K_{\rm{BPS}} \rho^{\gamma_{\rm{BPS}}} ,
\end{equation}
in which the parameter $K_{\rm{BPS}}$  and $\gamma_{\rm{BPS}}$ are determined by BPS EOS,
$P_{\rm{BPS}}\left(\rho\right)=K_{\rm{BPS}} \rho^{\gamma_{\rm{BPS}}} $.
When $\rho \geq \rho_{\rm{t}}$ we have
\begin{equation}
P_{2}\left(\rho \right)=K_{2} \rho^{\gamma} .
\end{equation}
At the transition density $\rho_{\rm{t}}$, $P_{1}\left(\rho_{\rm{t}} \right)=P_{2}\left(\rho_{\rm{t}} \right)$. Then we have
\begin{equation}
K_{2}=K_{\rm{BPS}} \rho_{\rm{t}}^{\gamma_{\rm{BPS}}}/ \rho_{\rm{t}}^{\gamma}.
\end{equation}

We use a new smooth curve to connect the two pressure lines with different exponent $\gamma_{\rm{BPS}}$ and $\gamma$,
\begin{equation}
P_{\rm{ns}}\left(\rho \right)=A_0\left[\left(\frac{\rho}{\rho_{\rm{t}}}\right)^{{\gamma_{\rm{BPS}}}{\alpha}}+\left(\frac{\rho}{\rho_{\rm{t}}}\right)^{{\gamma}{\alpha}}\right]^{\frac{1}{\alpha}},
\label{ns_eos}
\end{equation}
where
\begin{equation}
A_0=K_{\rm{BPS}} {\rho_{\rm{t}}}^{\gamma_{\rm{BPS}}}.
\end{equation}
A larger $\alpha$ means the new curve closer to the origin two lines (The BPS EOS and its extension before $\rho_{\rm{t}}$ as well as the stiffer polytropic model EOS). In this letter, $\alpha$ is set as $\alpha=1$ to obtain a smooth curve.
There are two free parameters in this model, the transition density $\rho_{\rm{t}}$ and the polytrope of exponent $\gamma$.
\subsection {Self-Bound Object on Surface}
Strange star (strange quark star and strangeon star) as a self-bound surface dense matter system,
 has non-zero surface density. We use the simple  polytropic
model with the pressure-density form,
\begin{equation}
P_{\rm{ss}}=K_{\rm{ss}} \rho^{\gamma} .
\end{equation}
When $\gamma=1$, this expression simplified to bag model (Alcock et al.\citealt{Alcock1986}).  In this case, with the linear EOS, the bag constant $B$ corresponds to the surface energy, $\rho_{\rm{s}}=4B$. Self-bound star has a sharp surface, that the pressure and density will decrease to zero in the fermi scale, which will not affect the mass-radius relation. We also involve the possibility that the strange star is  enveloped in thin nuclear crusts (e.g., Weber et al.\citealt{Weber2012}, Kettner et al.\citealt{Kettner1995}, Huang \& Lu\citealt{Huang1997}, Madsen\citealt{Madsen1999}, Weber et al.\citealt{Weber1994}, Xu\citealt{Xu2003}).

The pressure coefficient $K_{\rm{ss}}$ is determined by the sound speed.
Ab initio calculations could constrain the sound speed up to 1$\sim$2 $n_0$, but higher densities remain unconstrained (Tews et al.\citealt{Tews2018}).
Polytropic model may result in the superluminal problem, due to the sound speed
$v_s=\sqrt{\partial P/\partial \rho}$ monotonous increase with density, however Lu  et al.\cite{Lu2018} have proved that the actual propagation speed $v_{\rm{signal}} < c$ is always satisfied without destroying the causality.
We assume the stars are relativistic fluid and the sound speed square $v_s^2$ equals the conformal limit $1/3$ at the star surface which is assumed at high density,
\begin{equation}
v_s^2=\frac{dP}{d\rho}=K_{\rm{ss}}\gamma\rho_{\rm{s}}^{\gamma-1}=\frac{1}{3}.
\end{equation}
The pressure-density relation becomes
\begin{equation}
P_{\rm{ss}}\left(\rho \right)=\frac{1}{3\gamma\rho_{\rm{s}}^{\gamma-1}} \rho^{\gamma} ,
\label{ss_eos}
\end{equation}
with two free parameters, the surface density $\rho_{\rm{s}}$ and the polytropic exponent $\gamma$.
 \begin{figure}[hptb]
    \centering
    \includegraphics[viewport=10 10 580 580, width=8 cm,clip]{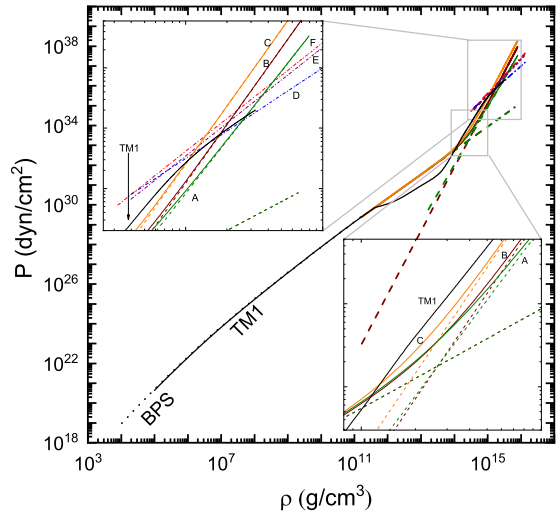}
    \caption{The EOS for gravity-bound and self-bound star. Solid lines A ($\rho_{\rm{t}}/ \rho_0$, $\gamma$)=(1.0, 2.9), B ($\rho_{\rm{t}}/ \rho_0$, $\gamma$)=(1.0, 3.2) and C ($\rho_{\rm{t}}/ \rho_0$, $\gamma$)=(0.7, 3.2)  are neutron star EOS, Dash-dot lines D ($\rho_{\rm{s}}/ \rho_0$, $\gamma$)=(1.5, 1.5), E ($\rho_{\rm{s}}/ \rho_0$, $\gamma$)=(1.5, 1.8) and F ($\rho_{\rm{s}}/ \rho_0$, $\gamma$)=(1.2, 1.8)  are self-bound star EOS. The dash lines indicate the corresponding two origin pressure-density relations for EOS A, B, and C. The thin solid line is the TM1 EOS. The dot line is the BPS EOS.}
    \label{eos}
\end{figure}
 \section{The Results}
 The EOS discussed above are shown in Figure~\ref{eos}. Logarithmic coordinates are used for the abscissa and ordinate, and in that case the curve slope equals the exponent $\gamma$ with density $\rho$. We use this nature to extend the BPS EOS to the transition density $\rho_{\rm{t}}$ for neutron star.
Solid lines A ($\rho_{\rm{t}}/ \rho_0$, $\gamma$)=(1.0, 2.9), B ($\rho_{\rm{t}}/ \rho_0$, $\gamma$)=(1.0, 3.2) and C ($\rho_{\rm{t}}/ \rho_0$, $\gamma$)=(0.7, 3.2) connect with BPS EOS are neutron star EOSs, while D ($\rho_{\rm{s}}/ \rho_0$, $\gamma$)=(1.5, 1.5), E ($\rho_{\rm{s}}/ \rho_0$, $\gamma$)=(1.5, 1.8) and F ($\rho_{\rm{s}}/ \rho_0$, $\gamma$)=(1.2, 1.8)  are self-bound star EOSs with a non-zero surface and pressure. Among these EOSs, A (D) and B (E) have the same transition density $\rho_{\rm{t}}$ (surface density $\rho_{\rm{s}}$), while B (E) and C (F) have the same polytropic exponent $\gamma$.  The polytropic exponent $\gamma$ is the symbol of EOS stiffness. The gravity-bound star EOSs we used has larger polytropic exponent $\gamma$ compared with the self-bound star EOSs. The RMF theory with TM1 parameter set provides excellent results for the properties of heavy nuclei ground states and 2.18 $M_\odot$ neutron star (Sugahara \& Toki\citealt{Sugahara1994}, Shen et al.\citealt{Shen2011}). We draw the TM1 EOS in Figure~\ref{eos} as a comparison. 
We found that after neutron drip out density there exits a slope decrease, which is considered as one of the reason that TM1 cannot support more massive neutron stars.

\begin{figure*}[hptb]
    \centering
    \includegraphics[viewport=5 5 580 580, width=8 cm,clip]{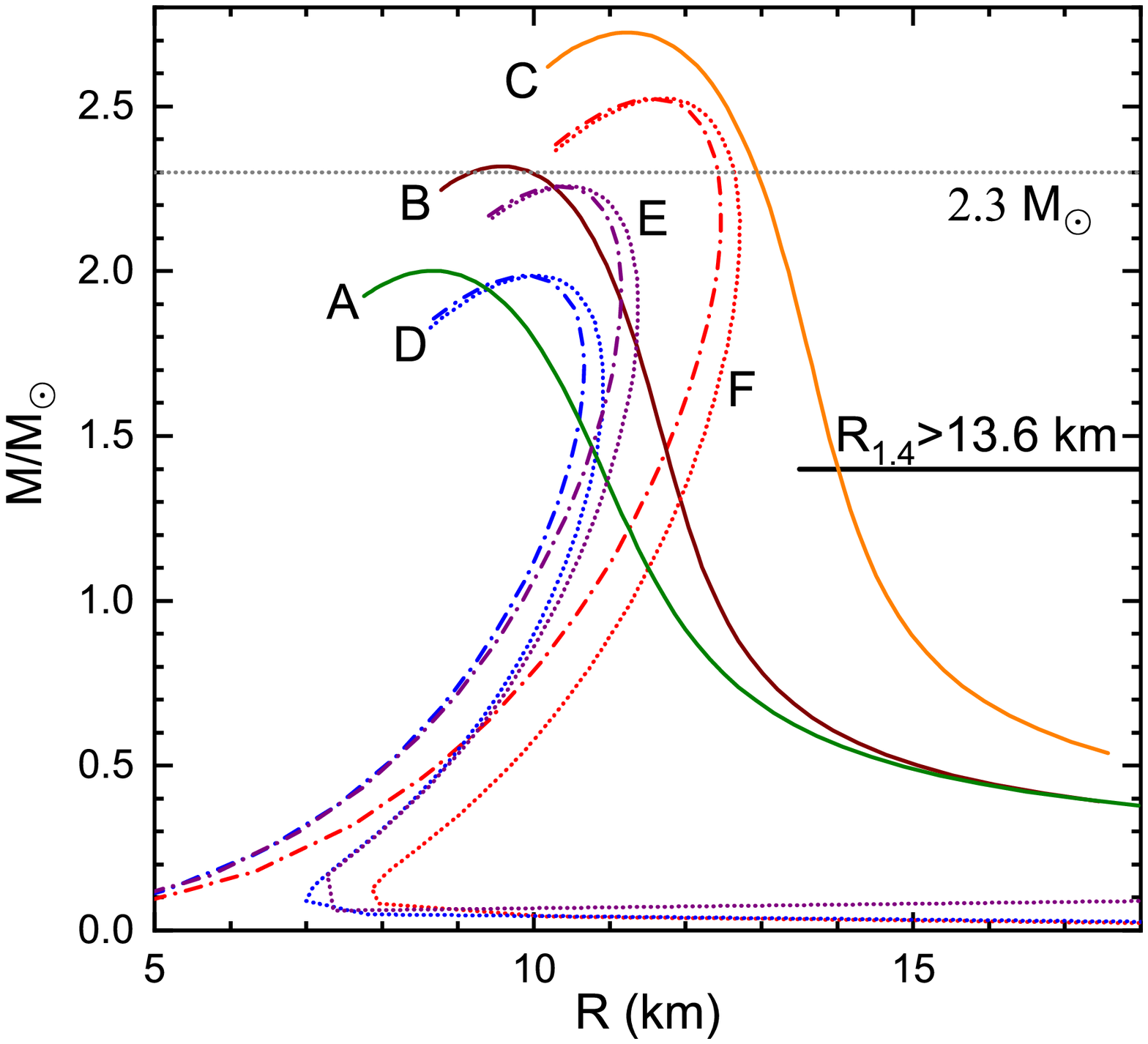}
    \includegraphics[viewport=5 5 580 580, width=8 cm,clip]{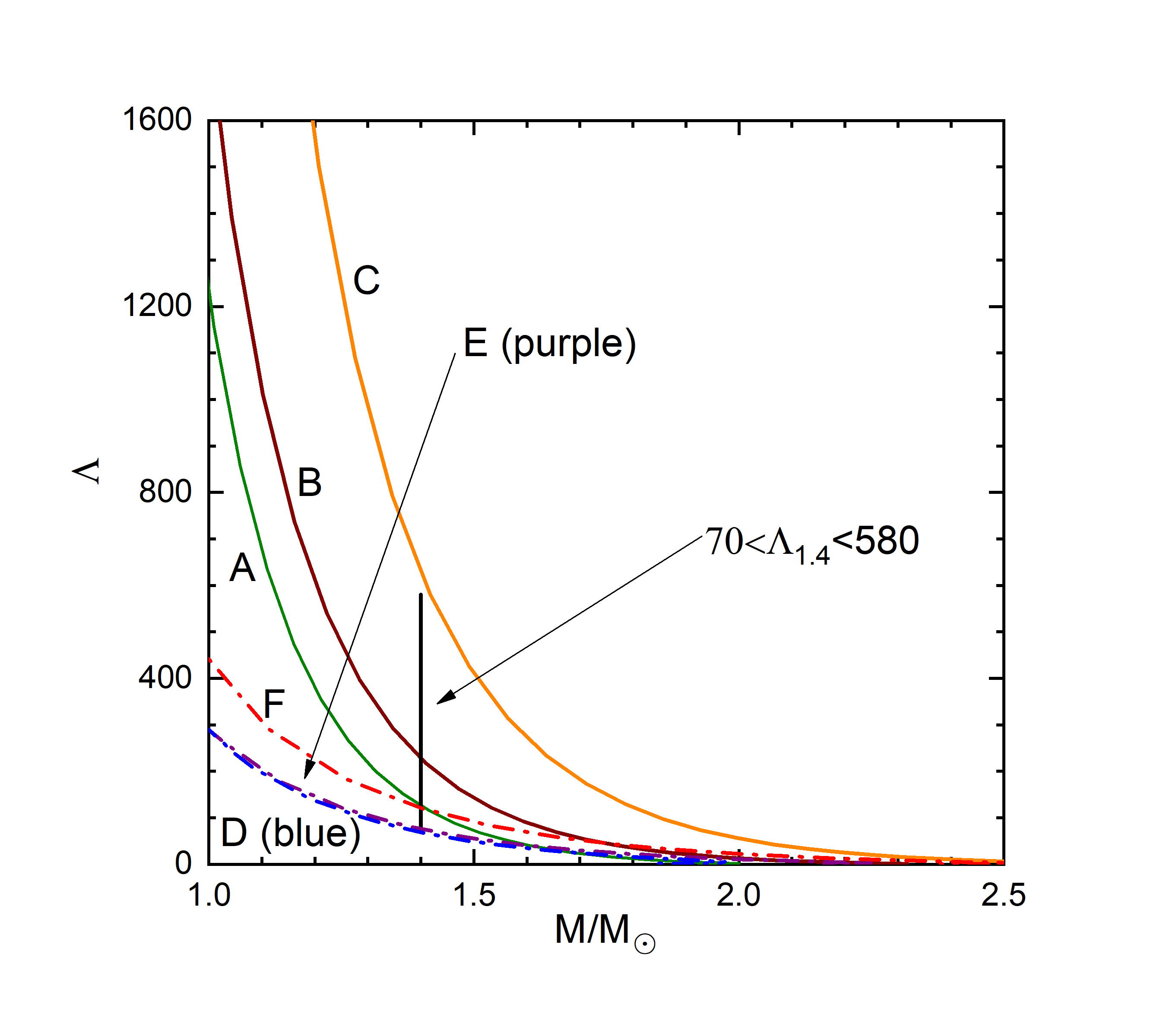}
    \caption{The mass-radius relation (left) and mass-tidal deformabilities relation (right) for gravity-bound and self-bound stars. Solid lines A ($\rho_{\rm{t}}/ \rho_0$, $\gamma$)=(1.0, 2.9), B ($\rho_{\rm{t}}/ \rho_0$, $\gamma$)=(1.0, 3.2) and C ($\rho_{\rm{t}}/ \rho_0$, $\gamma$)=(0.7, 3.2) are neutron star results, Dash-dot lines D ($\rho_{\rm{s}}/ \rho_0$, $\gamma$)=(1.5, 1.5), E ($\rho_{\rm{s}}/ \rho_0$, $\gamma$)=(1.5, 1.8) and F ($\rho_{\rm{s}}/ \rho_0$, $\gamma$)=(1.2, 1.8) are self-bound star results. The dot lines are self-bound star with a crust.}
    \label{rm}
\end{figure*}

Neutron star EOS combines the BPS EOS and a smooth curve derived by two polytropic pressure-density relations with different polytrope of exponent $\gamma$, in which the low-density pressure polytropic exponent $\gamma$ is same as the BPS EOS.
Inserting the EOS given by Eq.~\ref{ns_eos} and \ref{ss_eos} into Tolman-Oppenheimer-Volkoff (TOV) equation, we have the mass-radius relations for gravity-bound and self-bound star, see Figure~\ref{rm}. We also give the tidal deformalbilities as a function of neutron star mass in Figure~\ref{rm}.
A smaller transition density also means  large radius corresponds  to a relatively large mass because the hard core part could extend to lower density. The transition density $\rho_{\rm{t}} > \rho_0$ gives similar results with other effective models (e.g. HKP020, QMF18, SLy9, DD2, DDME2, NL3-$\omega\rho$, Haensel\citealt{Haensel1981}, Zhu et al.\citealt{Zhu2018}).
Stiffer EOS gives larger maximum mass and larger tidal deformalbilities. But strange star tidal deformalbilities are smaller than normal neutron star with similar maximum mass.
Line C ($\rho_{\rm{t}}/ \rho_0$, $\gamma$)=(0.7, 3.2) exceeds the $\Lambda_{1.4}$ range and breaks the $R_{1.4}$ restrict. Line D has $\Lambda_{1.4} \simeq 68$ which slightly exceeds the $\Lambda_{1.4}$ constraint.
Strange quark star or strangeon star may have a crust  which have a maximum density less than the neutron drip density. So that the Coulomb repulsive force can avoid the crust be absorbed by the strange star. We noticed that the mass-radius relation become normal neutron star like after adding the crust, since the crust is gravity-bound by the strange star core. Besides, the crust mass is so small compared to the self-bound star core that it negligibly increase the star mass but enlarge the radius when the self-bound star core is not massive enough.
Gravity-bound star model A (B) and self-bound star model D (E) have very similar maximum mass, while neutron star model polytropic exponent $\gamma$ is larger. This implies a gravity-bound system needs stiffer EOS than self-bound system to reach the same maximum mass.

 \begin{figure*}[htbp]
    \centering
    \includegraphics[viewport=5 5 480 480, width=8 cm,clip]{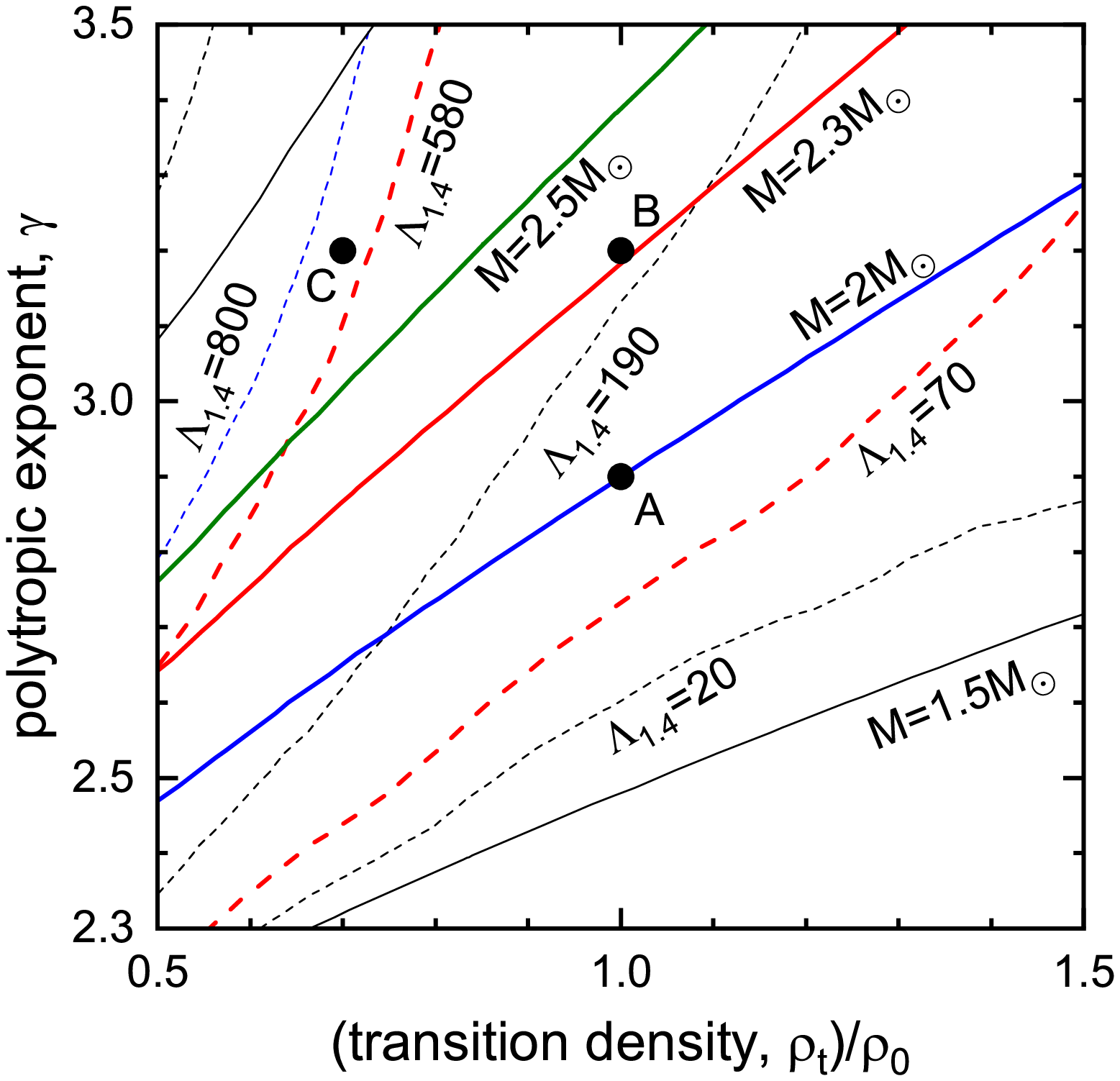}
    \includegraphics[viewport=5 5 480 480, width=8 cm,clip]{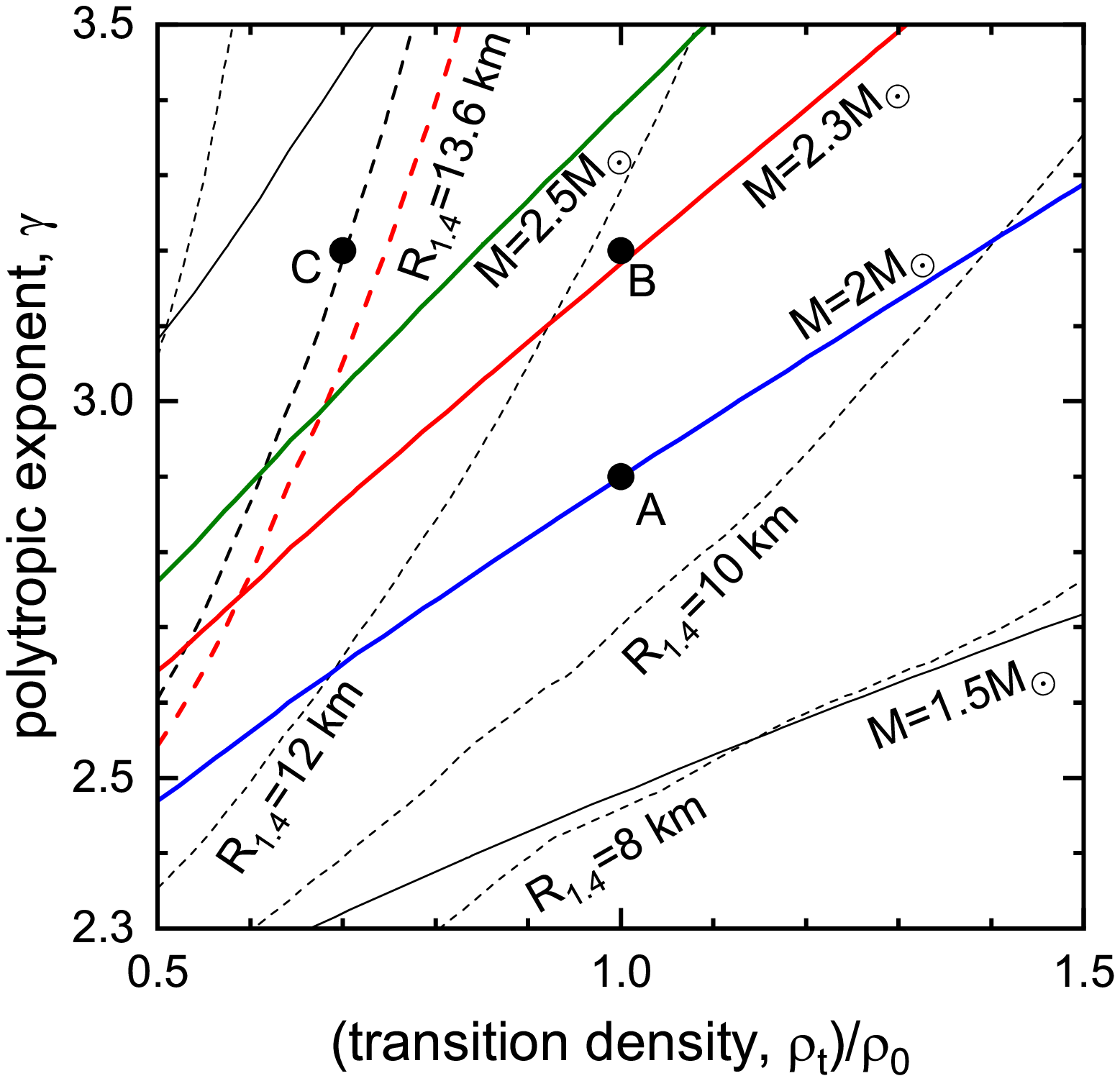}
    \caption{The normal neutron star $M_{\rm{max}}-\Lambda_{1.4}$ (left) and $M_{\rm{max}}-R_{1.4}$ (right) distribution in  $\rho_{\rm{t}}-\gamma$ parameter space. The solid lines show neutron star maximum mass, and the dash lines represent the 1.4 $M_\odot$ neutron star radius. Point A  and point B are  examples for 2 $M_{\odot}$ and  close to 2.3 $M_{\odot}$, respectively. Point C is ruled out by $\Lambda_{1.4}\leq580$ and $R_{1.4}\leq13.6$~km restricts.}
    \label{ns}
\end{figure*}
\begin{figure*}[hptb]
    \centering
    \includegraphics[viewport=5 5 480 480, width=8 cm,clip]{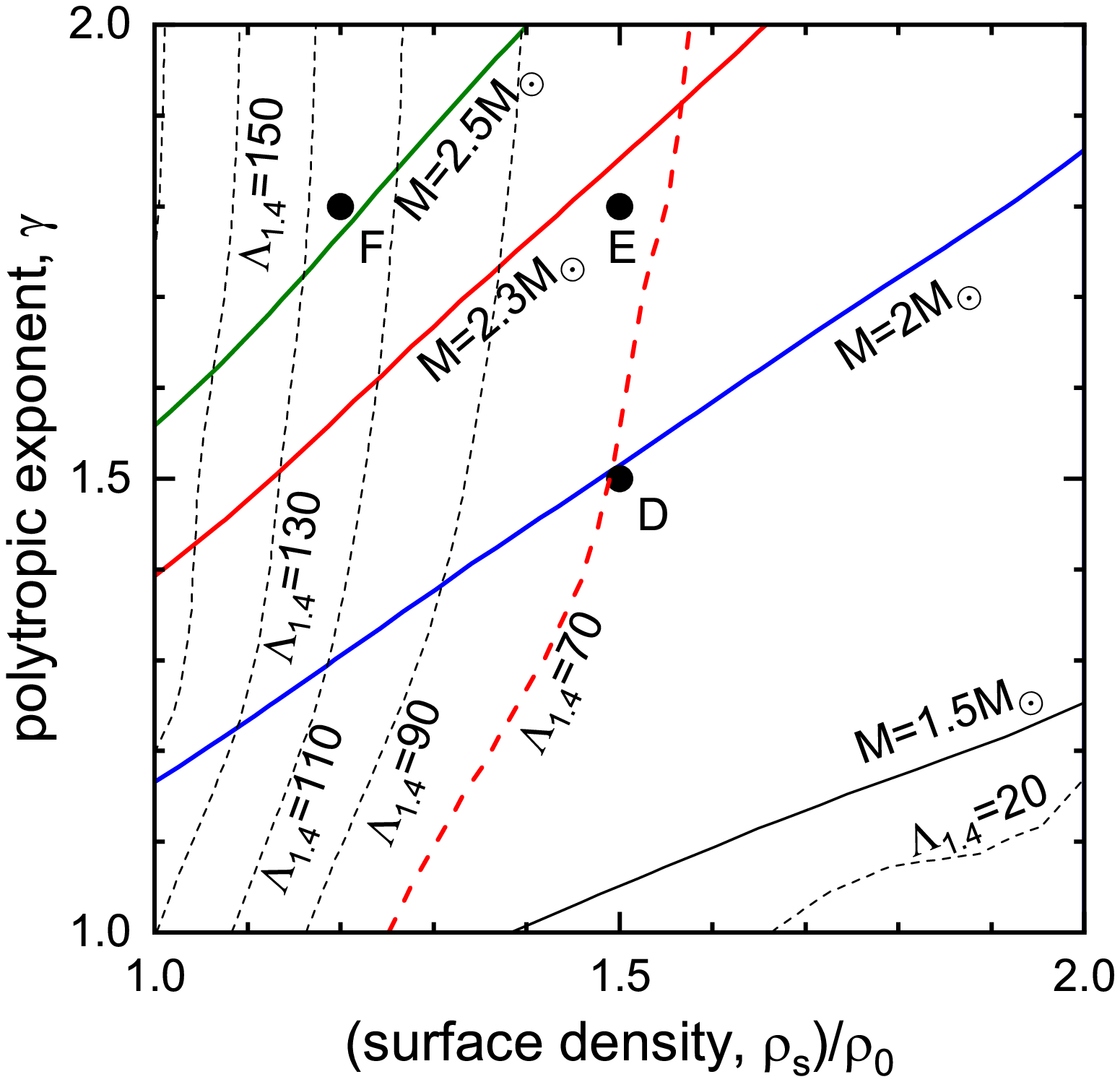}
    \includegraphics[viewport=5 5 480 480, width=8 cm,clip]{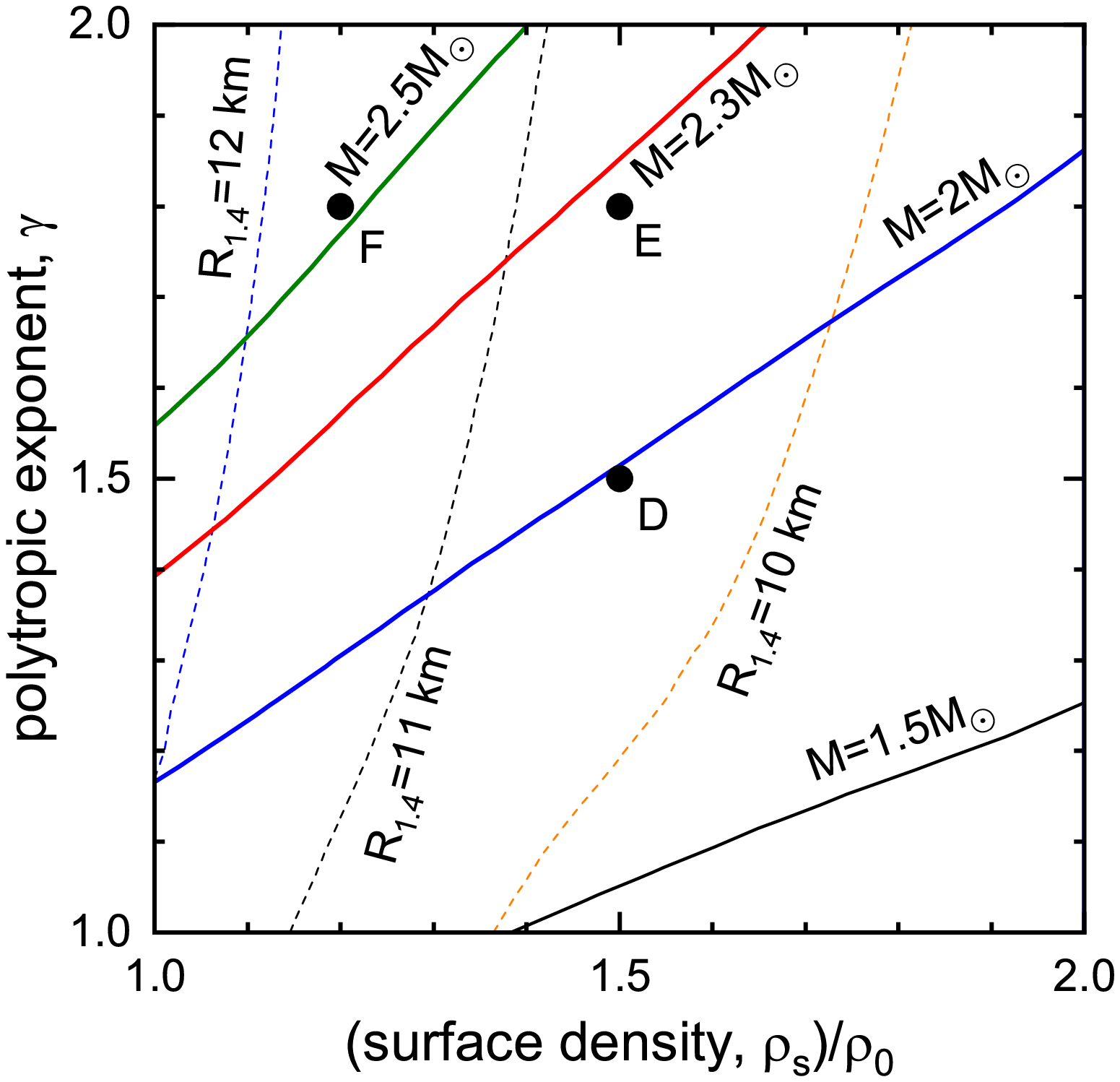}
    \caption{The strange star $M_{\rm{max}}-\Lambda_{1.4}$ (left) and $M_{\rm{max}}-R_{1.4}$ (right) distribution in $\rho_{\rm{s}}-\gamma$ parameter space. The solid lines indicate  strange  star maximum mass, and the dash lines represent the 1.4 $M_\odot$ strange  star radius. Point D, E and F are  examples close to 2 $M_{\odot}$, 2.3 $M_{\odot}$ and 2.5 $M_{\odot}$ self-bound star, respectively. Point D slightly beyond $70\leq \Lambda_{1.4}\leq 580$ restrict (see also Fig.~\ref{rm}).}
    \label{ss}
\end{figure*}

Figure~\ref{ns} show the mass and 1.4 $M_{\odot}$ neutron star tidal deformalbilities (radius) distribution in the transition density and polytropic exponent ($\rho_{\rm{t}}-\gamma$) parameter space (which can extend to higher values). Points A and B have the same transition density  $\rho_t=1.0\,\rho_0$, while points B and C have the same polytropic exponent $\gamma=3.2$.
A narrow area starts from ($\rho_{\rm{t}}/ \rho_0$, $\gamma$)=(0.50, 2.65) is available for $M\geq 2.3~M_\odot$ and $\Lambda_{1.4} < 580$. This area can certainly extend to overstep this figure range. The strong constraint $M\geq 2~M_\odot$ provides smaller ($\rho_{\rm{t}}$, $\gamma$) scope.
From this figure, a larger polytropic exponent $\gamma$ is needed to reach neutron star maximum mass limit, and the small transition densities $\rho_{\rm{t}}$ are restricted by the 1.4 $M_\odot$ tidal deformalbility.
Compared with these two distributions, it is found that radius range $9.8$ km $< R_{1.4} <$ $13.8$ km roughly consist with $70 \leq \Lambda_{1.4} \leq 580$.

Strange star usually use the surface density $(\rho_{s}/\rho_0)$=$(1.0\sim2.0)$. Figure~\ref{ss} is the similar distribution for self-bound object. Points D and E have the same surface density 1.5 $\rho_0$, while points E and F have the same polytrope of exponent $\gamma=1.8$. From our calculation, all the parameter range $(\rho_{s}/\rho_0, \gamma)$=$(1.0\sim2.0, 1.0\sim2.0)$ satisfy the $R_{1.4}\leq13.6$ km limit. But $\Lambda_{1.4} \geq 70$ exclude larger surface density that  $\rho_{\rm{s}}<1.58$ if $\gamma<2.0$. $2~M_\odot$ self-bound system needs $\gamma>1.18$, and $\gamma > 1.40$ for $2.3 M_\odot$ self-bound system. Self-bound star scene could have massive maximum mass and relatively small tidal deformability simultaneously.  $R_{1.4}<12.5$ is 
needed by $\rho_{\rm{s}}\geq 1.0$, and $R_{1.4}>10.5$ is roughly consistent with 
$\Lambda_{1.4} \geq 70$.

An evidence trend is observed that the pulsar maximum mass increase with larger polytropic exponent $\gamma$ and smaller transition density $\rho_{\rm{t}}$ (surface density $\rho_{\rm{s}}$) for gravity-bound scene (self-bound scene). 
An obvious phenomenon is found that the gravity-bound star EOS are stiffer than the self-bound star to attain similar neutron star maximum mass.

\begin{figure}[htbp]
	\centering
	\includegraphics[width=0.6\textwidth]{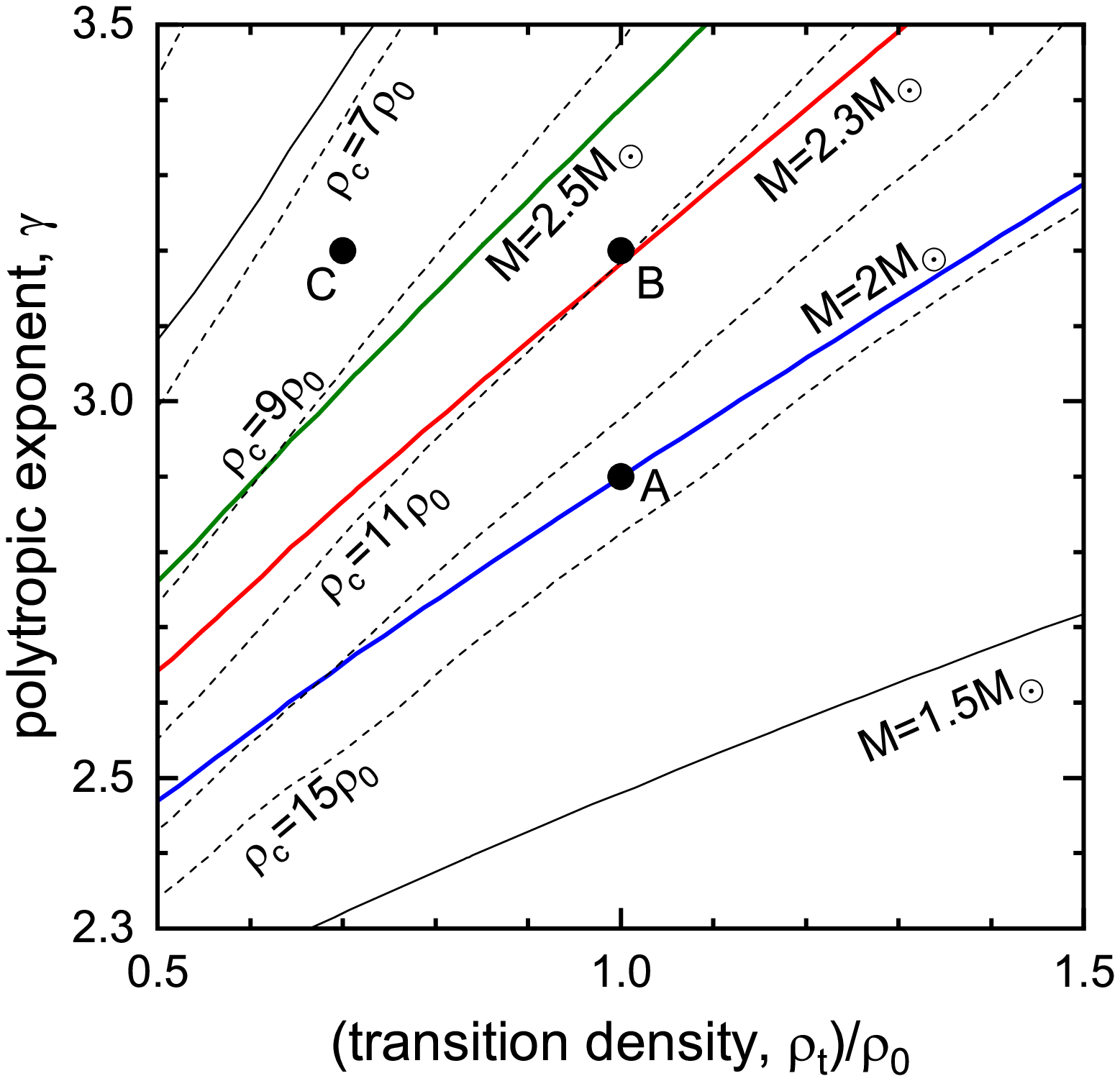}
	\caption{The normal neutron star maximum mass-central density ($M_{\rm{max}}-\rho_0$) distribution in $\rho_{\rm{s}}-\gamma$ parameter space. The solid lines show neutron star maximum mass, and the dash lines represent the center density. Point A  and B are  examples for 2 $M_{\odot}$ and  close to 2.3 $M_{\odot}$, respectively. }
	\label{ns-nc}
\end{figure}
\begin{figure}[hptb]
	\centering
	\includegraphics[width=0.6\textwidth]{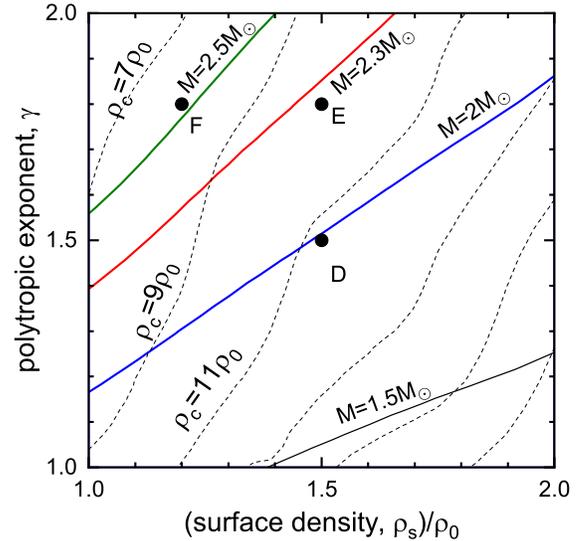}
	\caption{The self-bound star maximum mass-central density ($M_{\rm{max}}-\rho_0$) distribution in $\rho_{\rm{s}}-\gamma$ parameter space. The solid lines indicate  strange  star maximum mass, and the dash lines represent the center density with 2 $\rho_0$ spacing. Point D  and E are  examples close to 2 $M_{\odot}$ and  2.3 $M_{\odot}$ self-bound star, respectively.}
	\label{ss-nc}
\end{figure}

Figure~\ref{ns-nc} and Figure~\ref{ss-nc} are the mass and central density distribution in the $\rho_{\rm{t}}(\rho_{\rm{s}})-\gamma$ parameter space for normal neutron star and strange star respectively. 
Smaller central density always match with larger maximum mass and 
normal neutron star has larger central density than strange star with the same maximum mass. 
The deconfined quark matter is expected to appear at relative high density, beyond where quarks can no longer considered belonging to specific baryons. For a typical baryon radius $r_b=0.5$ fm, quark percolation could occur at $1/\sqrt{2}(4/3\pi r_b^3)/\rho_0$ $\approx 8.44~\rho_0$ (face center cubic). From which the polytropic exponent should have an alter value that quark matter construction replaced original hadronic matter. In other words, the polytropic model should have another segment.
Since the core part contribute most to the neutron star,  a smaller transition density (surface density) will lead to smaller central density with the same maximum mass.
\section{Conclusions and Discussions}
The existence of massive neutron star requires stiff core EOS. However, the tidal deformability restrict and supernova explosion prefer a soft normal neutron star crust. We apply the polytropic model to examine how stiff the neutron star EOS should be and the transition density (surface density) range.
We apply vanishing surface density for normal neutron star and non-zero surface density for strange star. As a self-bound system, strange star could have smaller radius and larger mass with the same polytropic exponent compared with normal neutron star.

A small transition density $\rho_{\rm{t}} > 0.50~\rho_0$ and a large polytropic exponent $\gamma > 2.65$ are beneficial to the $M_\textrm{max}>2.3~M_\odot$ conventional neutron star, while for $M_\textrm{max}>2.3~M_\odot$ strange star, the polytropic exponent $\gamma > 1.40$ is required to $\rho_{\rm{s}} \sim 1.0~\rho_0$. A small transition density ($\rho_{\rm{t}}/\rho_0 < 0.50$ for 2.3 $M_\odot$) may break the 1.4 $M_\odot$ tidal deformalbility restrict, while self-bound system model requires $\rho_{\rm{s}}/\rho_0 < 1.58$ for $\gamma < 2.0$. 
By comparing the $M_{\rm{max}}-\Lambda_{1.4}$ and $M_{\rm{max}}-R_{1.4}$ distribution we derive rough results of common neutron star radius range, which is $9.8~\rm{km} < R_{1.4} < 13.8~\rm{km}$ for normal neutron stars and $10.5~\rm{km} < R_{1.4} < 12.5~\rm{km}$ for strange stars.
With this work we transform the mass-radius constraints into the the stiffness transition density and the polytropic exponent (stiffness measurement parameter) parameter space. These are meaningful for other phenomenological model.

\acknowledgments
We acknowledge useful discussions at the pulsar group of PKU.
This work is supported by the National Key R$\&$D Program of China (Grant Nos. 2018YFA0404703, and 2017YFA0402602), the National Natural Science Foundation of China (Grant Nos. 11673002, and U1531243), and the Strategit Priority Research Program of CAS (Grant No. XDB23010200).

\end{document}